\documentclass[3p,times,procedia]{elsarticle}

\usepackage{ecrc}

\volume{00}

\firstpage{1}

\journalname{Physics Procedia}

\runauth{}

\jid{phpro}

\jnltitlelogo{Physics Procedia}

\CopyrightLine{2011}{Published by Elsevier Ltd.}

\usepackage{amssymb}
\usepackage{amsmath}

\usepackage[figuresright]{rotating}

\begin{document}

\begin{frontmatter}



\dochead{}

\title{Improving Dark Matter Searches by Measuring the Nucleon Axial Form Factor: Perspectives from MicroBooNE}


\author{Tia Miceli, Vassili Papavassiliou, Stephen Pate, Katherine Woodruff \newline for the MicroBooNE Collaboration}

\address{NMSU Department of Physics, MSC 3D P.O. Box 30001, Las Cruces, NM 88003}

\begin{abstract}
The MicroBooNE neutrino experiment at Fermilab is constructing a liquid-argon time-projection chamber for the Booster Neutrino Beam to study neutrino oscillations and interactions with nucleons and nuclei, starting in 2014.
We describe the experiment and focus on its unique abilities to measure cross sections at low values of $Q^2$.
In particular, the neutral-current elastic scattering cross section is especially interesting, as it is sensitive to the contribution of the strange sea quark spin to the angular-momentum of the nucleon, $\Delta s$.
Implications for dark-matter searches are discussed.
\end{abstract}


\end{frontmatter}



\section{Motivation}
\label{sec:motivation}

\begin{figure}[hbtp]
  \begin{center}
  \scalebox{0.2}{\includegraphics{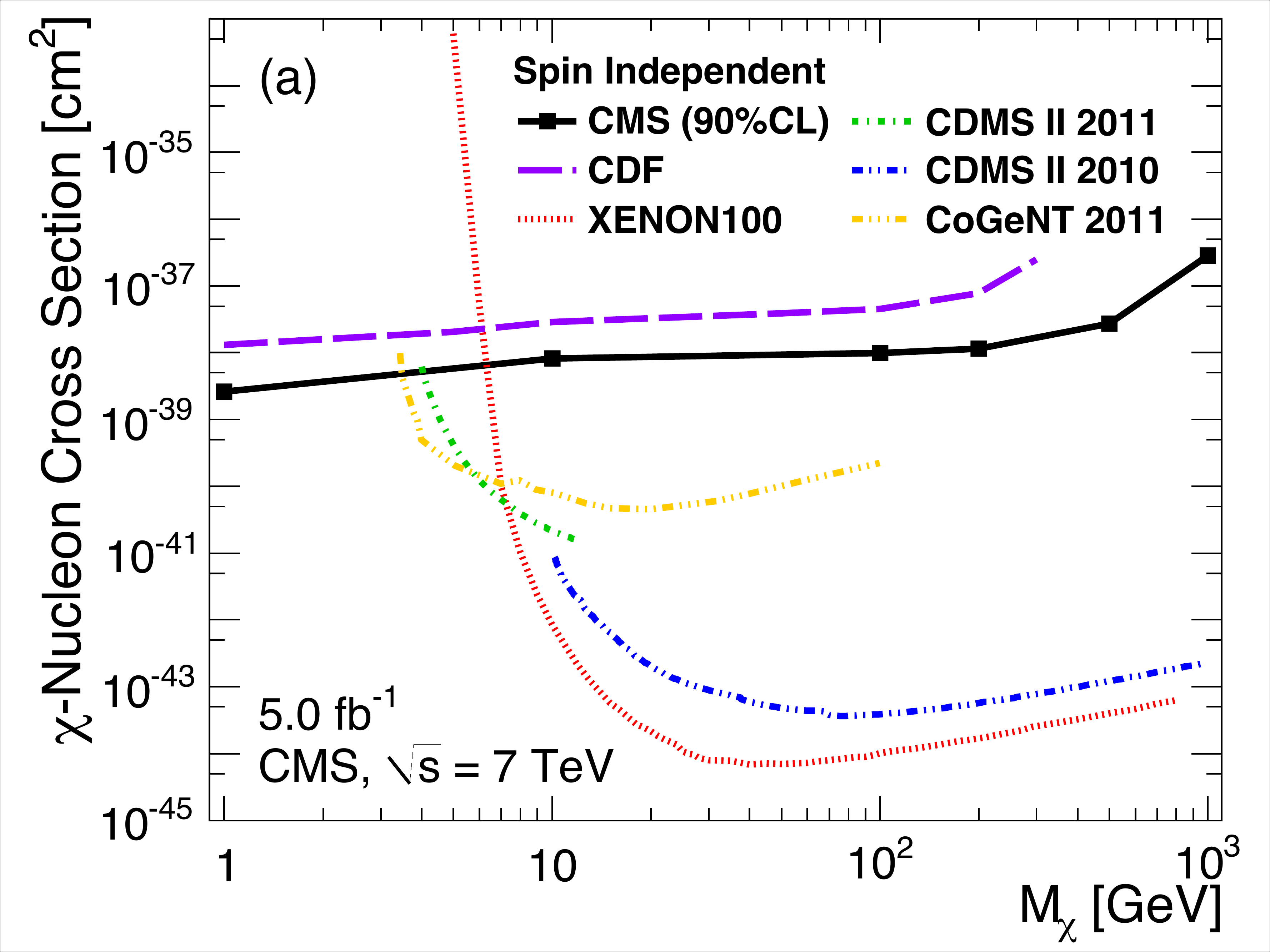}}
  \scalebox{0.2}{\includegraphics{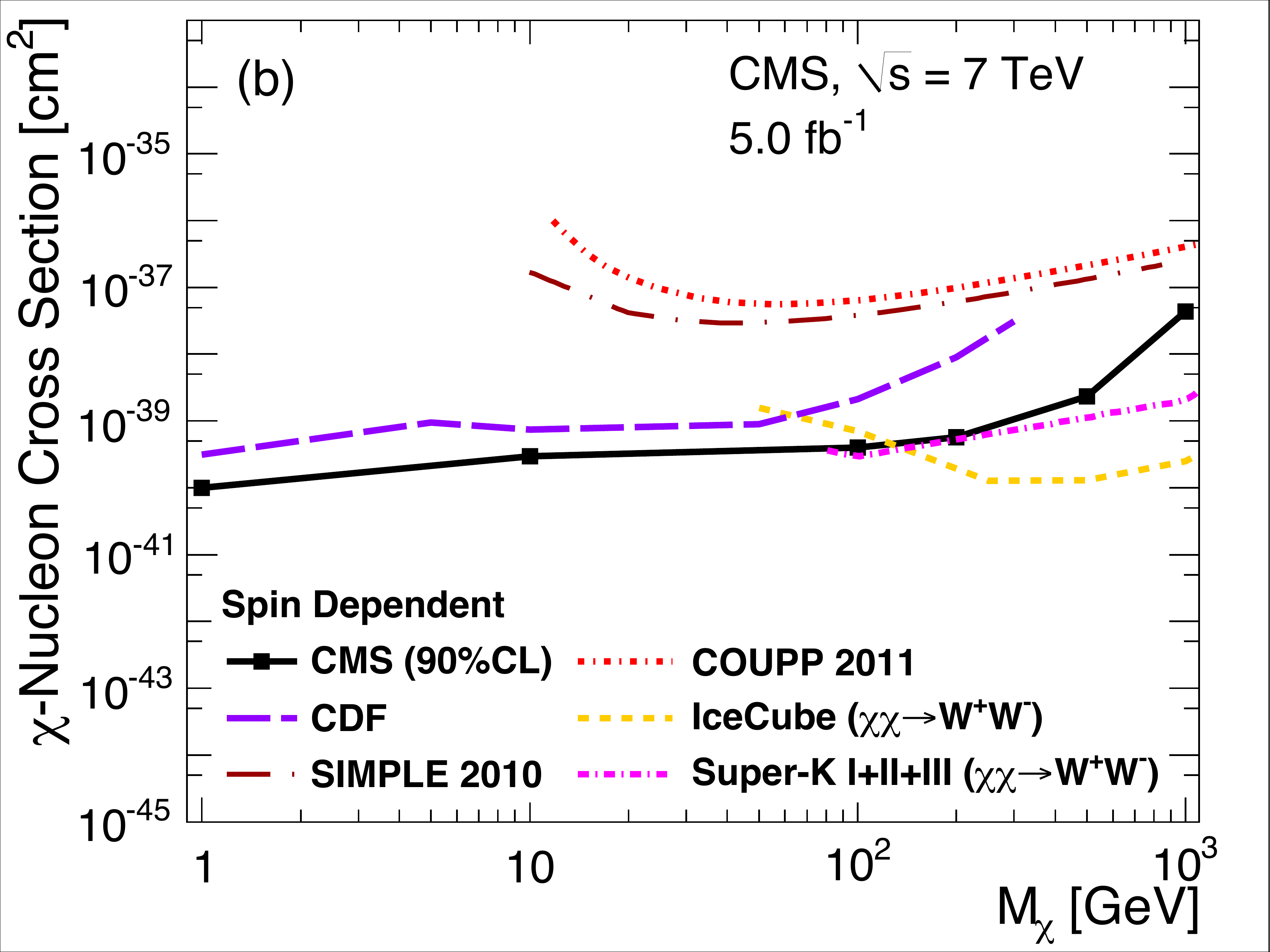}}
  \caption{Dark matter cross sections for spin-independent (a) and spin-dependent (b) dark matter.~\cite{cms}.}
  \label{fig:cs}
  \end{center}
\end{figure}

Astrophysics and cosmology estimates that dark matter makes up $26.8\%$ of the universe~\cite{planck}.
Particle physics experiments have yet to identify a dark matter candidate.
A variety of experiments ranging from nuclear recoil direct-detection experiments, to collider indirect-experiments have been searching for dark matter, and although they have not identified a candidate yet, they are whittling down the possible parameter space for such a particle. 
Current physical models allow dark matter cross sections to be dependent or independent of the spin of the scattering target.
The limits on these cross sections are shown in Fig.~\ref{fig:cs}.

\begin{figure}[hbtp]
  \begin{center}
  \scalebox{0.5}{\includegraphics{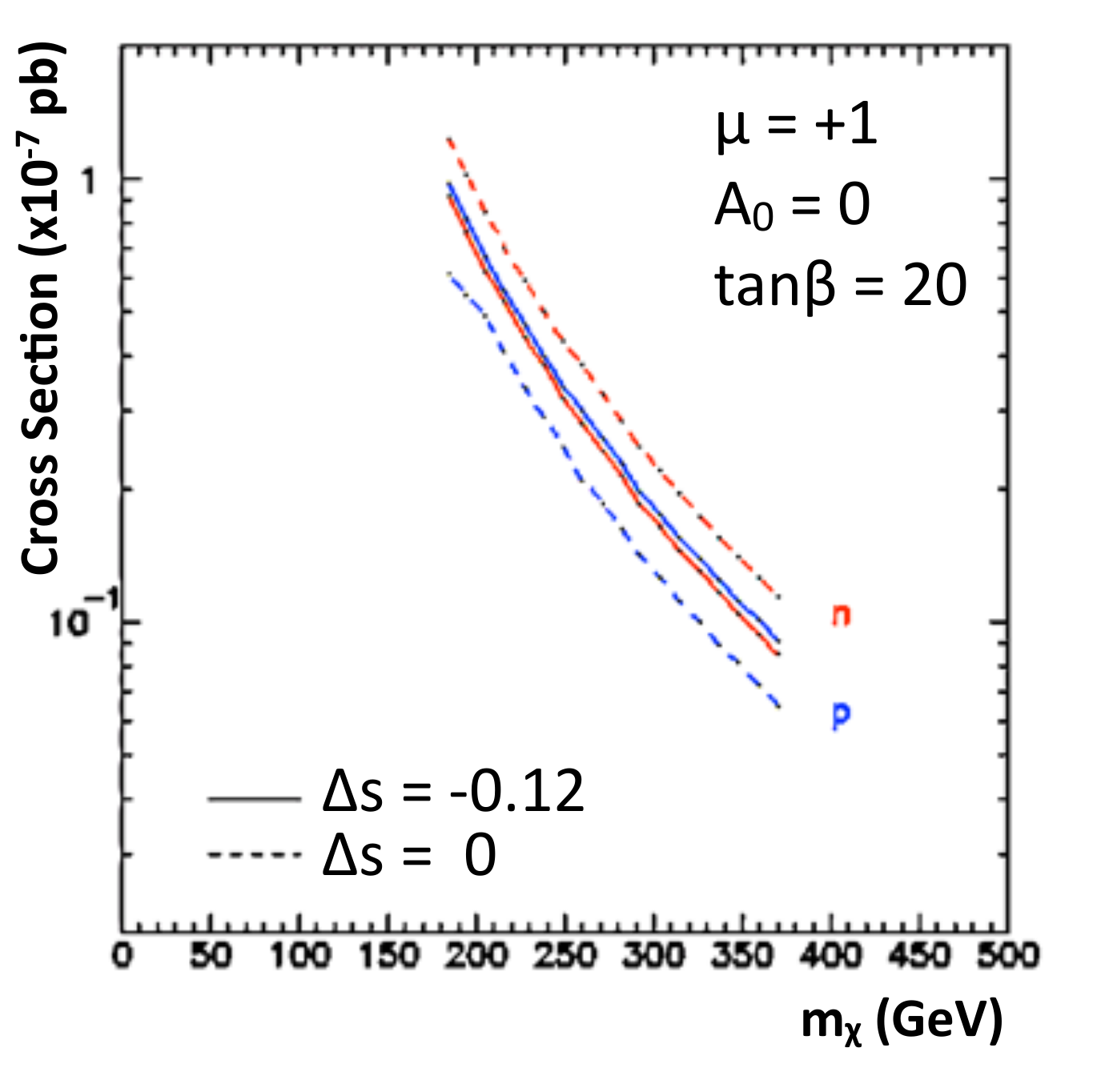}}
  \caption{Dark matter cross sections for targets with an odd number of protons (blue) or an odd number of neutrons (red) for negative values of $\Delta s$ (solid) or a zero value (dashed). Calculated using the DarkSUSY simulation tool assuming SUSY parameters of $\mu= +1$, $A_0=0$, and $\tan\beta = 20$ ~\cite{ds}\cite{Gondolo:2004sc}.}
  \label{fig:dmcs}
  \end{center}
\end{figure}

To reduce the uncertainty on the spin dependent cross section, it is important to know the spin structure of the target.
In the direct detection experiments, the target is a nucleus.
The spin dependent cross section will be zero unless there is an unpaired nucleon spin in the nucleus.
By improving our understanding of the nucleon spin structure, the uncertainties of the scattering target, and therefore the uncertainties on the dark matter cross section measurement will be improved.
The net helicity of strange and anti-strange quarks in a nucleon, $\Delta s$, is rather poorly known, and at the same time it has a strong influence on dark matter detection cross sections, as shown in Fig.~\ref{fig:dmcs}. The estimates in Figure 2 were made using the DarkSUSY simulation tool~\cite{ds}\cite{Gondolo:2004sc} assuming SUSY parameters of $\mu= +1$, $A_0=0$, and $\tan\beta = 20$.
If $\Delta s$ is negative, as suggested by polarized deep-inelastic scattering experiments, targets with an odd number of unpaired protons will see a higher cross section (solid blue), whereas targets with an odd number of unpaired neutrons will see a lower cross section (solid red).

These proceedings will summarize the current state of knowledge of the nucleon spin structure, and present a method to measure the strange component to the nucleon spin.
This is the variable which is most in need of improvement in order to better estimate dark matter cross sections.
We will highlight the improvement capabilities from the new MicroBooNE detector at Fermilab.

\section{Form Factor Global Fits}
\label{sec:ffgf}

Calculating the scattering amplitude of a particle with a nucleon is non-trivial since the nucleon is not point-like.
Experimentally determined form factors are used to parameterize the scattering matrix element into pieces.
Each form factor depends on the momentum transferred, $Q^2$.
The form factors for the electromagnetic current and charged weak current are well understood for $Q^2<1$ GeV$^2$.
However, the matrix element for the neutral weak current can be improved, specifically the axial form factor which is related to the spin structure of the nucleon.
The axial form factor has components dependent on the up, down, and strange quark composition of the nucleon, Eq.~\ref{eq:GA}
\begin{equation}\label{eq:GA}
G^{Z,p}_{A} = \frac{1}{2}(-G^u_A + G^d_A + G^s_A)
\end{equation}
The combination of form factors $-G_A^u+G_A^d$ is well determined from charged current neutrino reactions and elastic and deep inelastic electron and muon scattering experiments, as well as neutron decay experiments~\cite{SPate}.
The last form factor, $G^s_A$, approaches the value of the strange quark spin, $\Delta s$, as the momentum transfer squared, $Q^2$, goes to zero, Eq.~\ref{eq:GsA}.
\begin{equation}\label{eq:GsA}
G^s_A(Q^2 \rightarrow 0) = \Delta s
\end{equation}
This quantity can be measured in neutral current neutrino scattering experiments provided that the detector has detailed tracking capabilities.

\begin{figure}[hbtp]
  \begin{center}
  \scalebox{0.52}{\includegraphics{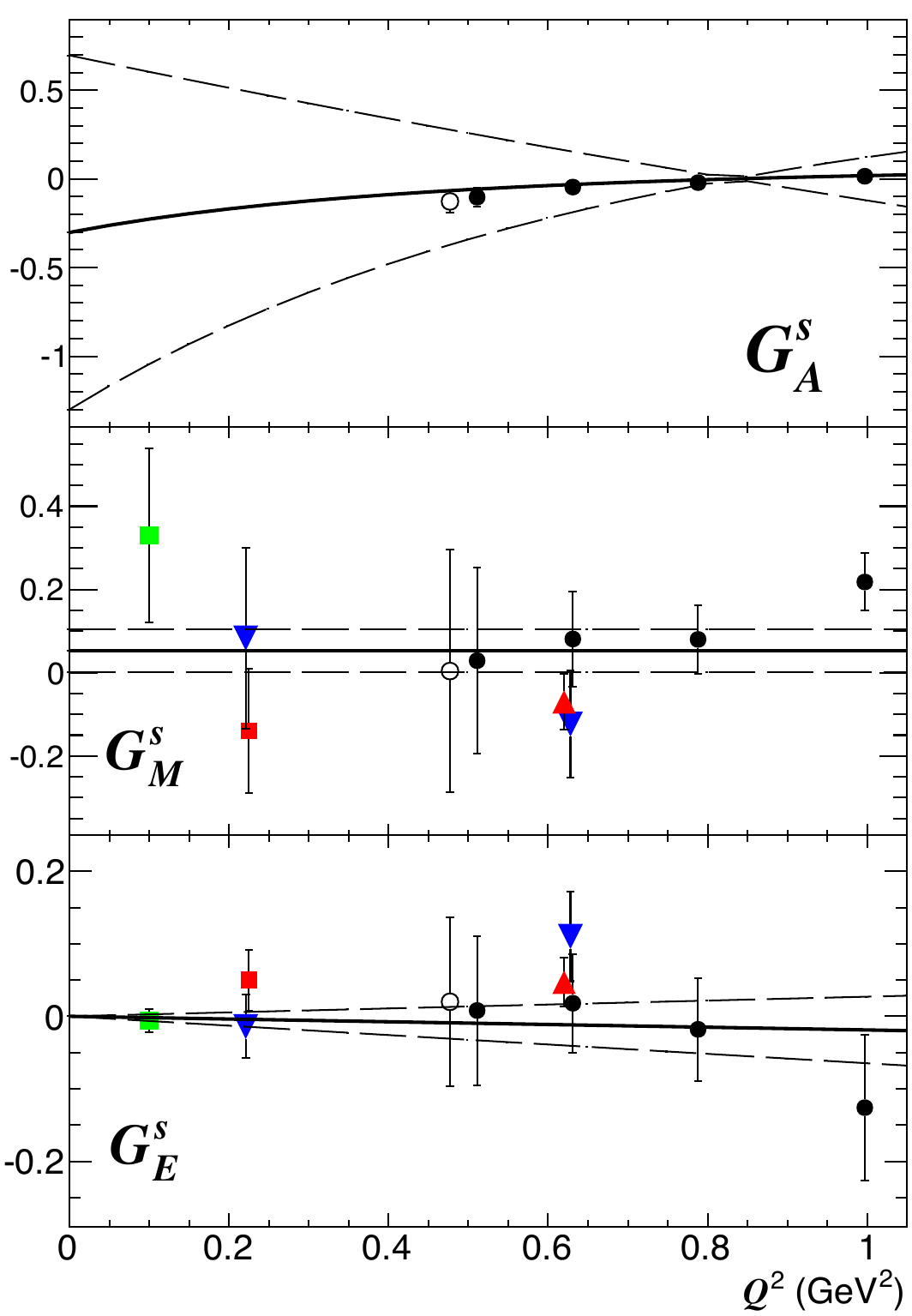}}
  \scalebox{0.34}{\includegraphics{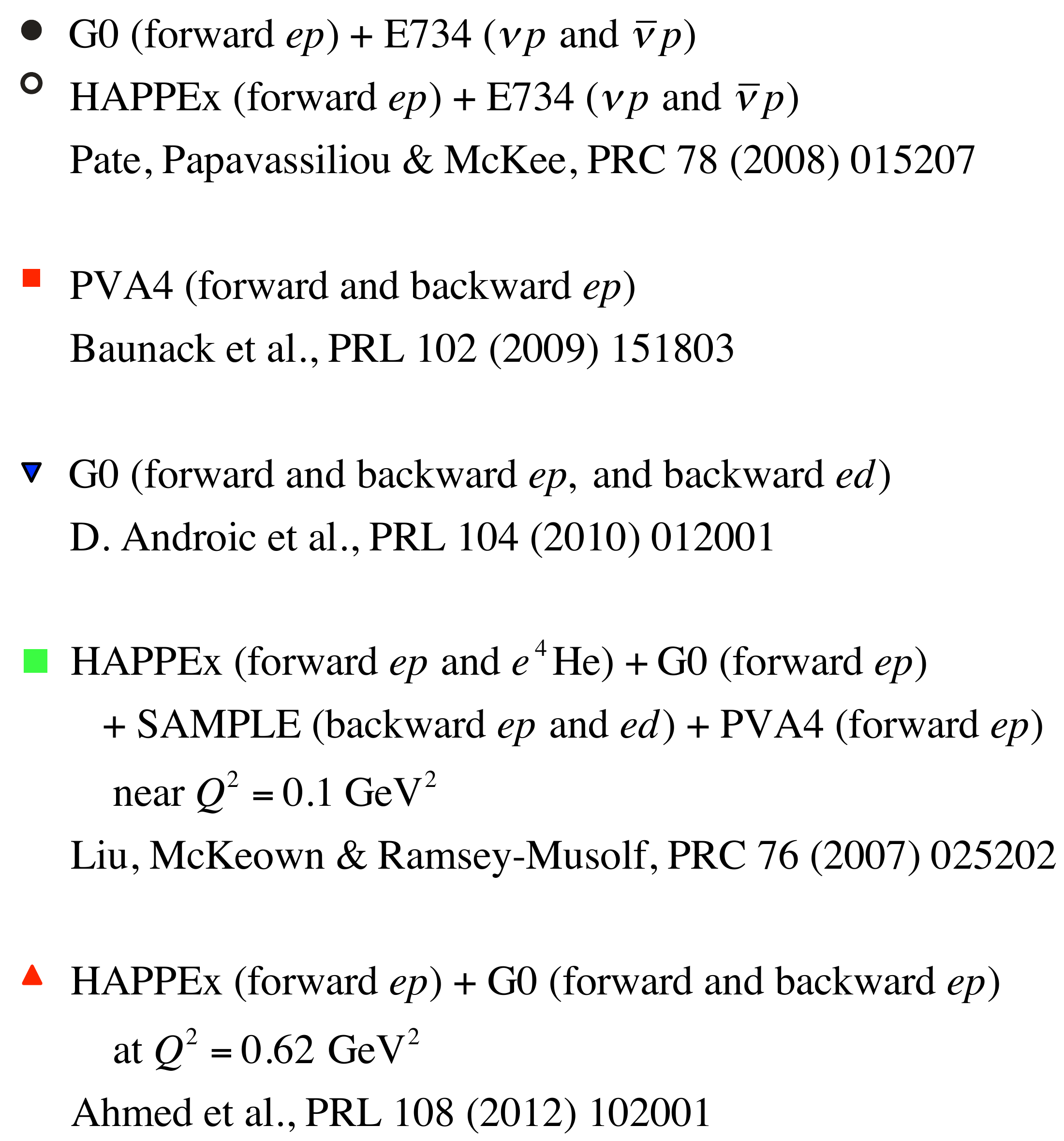}}
  \caption{Points: determinations of the strangeness form factors at specific $Q^2$ values, drawing on specific experiments.  Solid line:  Global fit of all 48 available data points.  Dashed line: 70\% confidence interval for the global fit.~\cite{SPate}~\cite{SPate2}.}
  \label{fig:globalfit}
  \end{center}
\end{figure}

A total of 48 data points from experiments of lepton elastic and quasi-eleastic electroweak scattering spanning from 1987 to present were simultaneously fit as a function of $Q^2$~\cite{SPate}~\cite{SPate2}.
Each of these experiments probes the neutral current form factors of Eq.~\ref{eq:GsA} shown in Fig.~\ref{fig:globalfit}.
The uncertainty on $G^s_A$ widens as $Q^2 \rightarrow 0$ due to the lack of neutrino data in this region.
The MicroBooNE experiment can measure $G^s_A$ at low $Q^2$ by measuring neutral current quasi-elastic scattering of neutrinos off  argon atoms.

\section{Current Experimental Results}
\label{sec:current}

The BNL experiment 734 was a layered detector that measured neutrino and anti-neutrino cross sections.
By taking the ratio of $\nu p$ and $\overline{\nu} p$ elastic scattering cross sections, and fitting the results to a functional form for $G^s_A$, a wide range of $\Delta s$ were found to describe the data well~\cite{garvey}.
The values of $\Delta s$ ranged from $0$ to $-0.21\pm0.10$.
MicroBooNE will be capable of measuring lower $Q^2$, thereby narrowing down the value of $\Delta s$.

\begin{figure}[hbtp]
  \begin{center}
  \scalebox{0.6}{\includegraphics{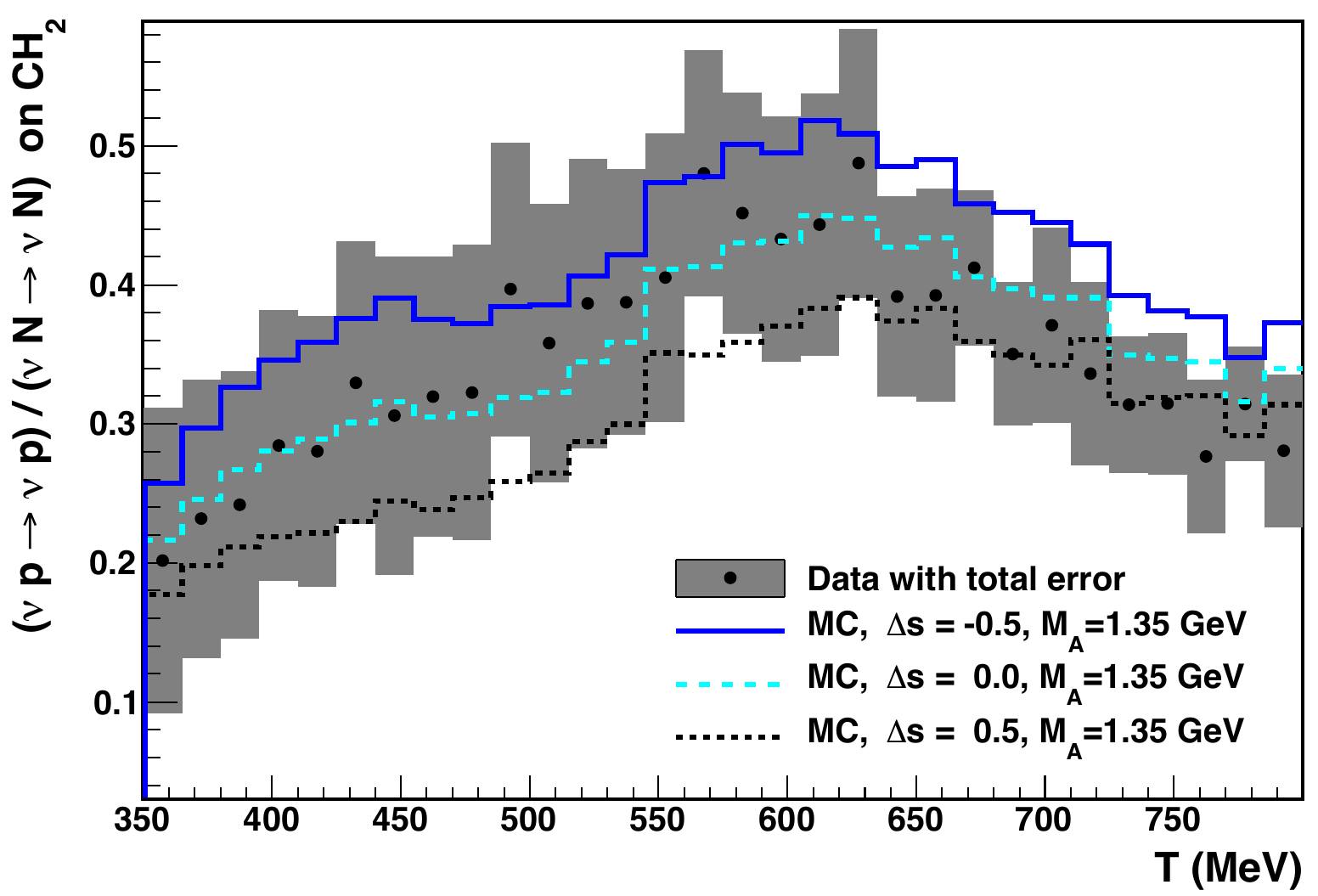}}
  \caption{Ratio of neutral current proton scattering to all neutral current scattering in the MiniBooNE detector. Black dots and grey bands are data, and uncertainty of the data, solid blue line is the simulated ratio assuming $\Delta s$ is -0.5, dashed cyan for $\Delta s$ is 0, and dashed black for $\Delta s$ is 0.5\cite{miniboone}.}
  \label{fig:miniboone}
  \end{center}
\end{figure}

The MiniBooNE experiment, an oil chernkov neutrino detector, also had limited sensitivity to the measurement of $\Delta s$.
By measuring the ratio of cross sections of neutral current scattering off a proton to the neutral current scattering off all nucleons, MiniBooNE could explore it's sensitivity to $\Delta s$.
Fig.~\ref{fig:miniboone} shows how this ratio varies with the kinetic energy of the nucleon, $T$~\cite{miniboone}.
Three hypotheses for $\Delta s$ were simulated and the cross section ratio data plotted with the data.
The simulated set which most accurately describes the data, provides the measurement of $\Delta s$, $0.08\pm0.26$.
The wide uncertainty on this result is due to the low energy limitation of MiniBooNE.
MicroBooNE will be able to measure lower energy protons, down to at least $40$~MeV, thereby being sensitive to even lower values of $Q^2$ than previous experiments. 

\section{Anticipated improvement with MicroBooNE data}
\label{sec:experiment}

The MicroBooNE detector, which is now under construction at Fermilab, will be well-suited for the measurement of $G^s_A$ at low $Q^2$.
The ratio of neutral current (NC) elastic cross section to charged current (CC) quasi-elastic cross section is sensitive to $\Delta s$.
By using a ratio of cross sections (Eq.~\ref{eq:ratio}), some nuclear, detector, and flux effects common to both neutral current and charged current interactions will be cancelled.
\begin{equation}\label{eq:ratio}
\text{ratio}(\Delta s)= \frac{ \sigma(NC)^{\text{elastic}} }{ \sigma(CC)^{\text{quasi-elastic}} } =\frac{\sigma( \nu p \rightarrow \nu p )}{\sigma( \nu n \rightarrow \mu p )}
\end{equation}

By fitting a function to this ratio, the value of $\Delta s$ can be extracted, similar to the procedure described for MiniBooNE.
MicroBooNE's improved particle identification enables measurements down to a lower $Q^2$ than MiniBooNE.
At MicroBooNE, a 40~MeV proton will provide the minimum detectible track length (spreading over 5 detection wires).
This enables a measurement of $Q^2$ down to $\sim 0.08$ GeV$^2$, which will be the lowest $Q^2$ neutral current measurement for any neutrino scattering experiment.

A simple simulation of 50k $\nu$-Ar events were generated using Nuance v3, this corresponds to approximately $2\times10^{20}$ protons on target (approximately a year of running).
There were 22209 charged current scatters on a neutron target, 4343 neutral current scatters on a neutron target, and 2760 neutral current scatters on a proton target.
The mean energy of the beam was $E_\nu \sim 1$~GeV, simulating what can be expected from the Booster Neutrino Beam at Fermilab.

\begin{figure}[hbtp]
  \begin{center}
  \scalebox{0.55}{\includegraphics{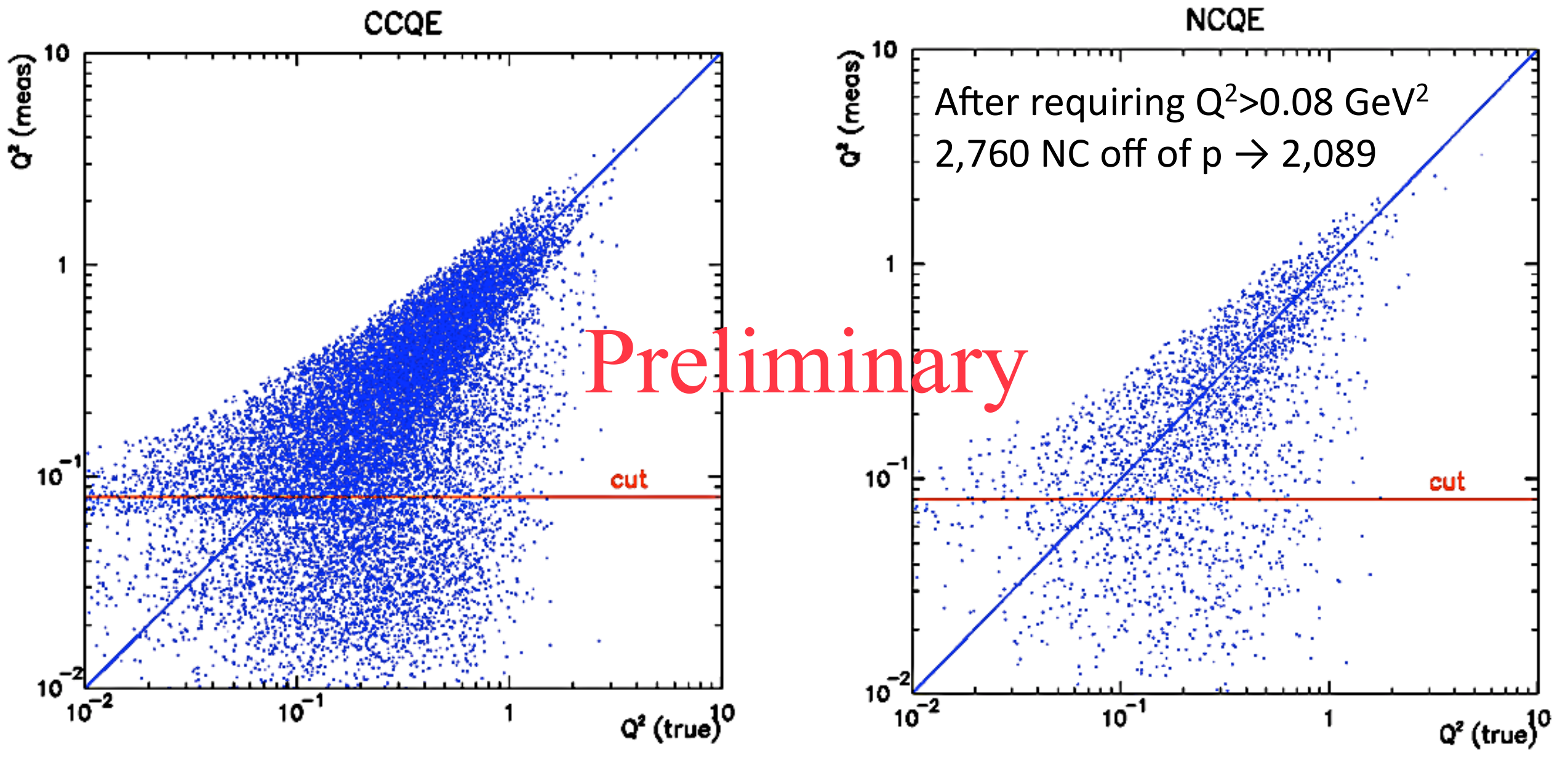}}
  \caption{Measured momentum transfer squared versus true momentum transfer squared for charged current quasi-elastic scatters and neutral current quasi-elastic scatters. The cut shows where the minimum measurable $Q^2$ is, $\sim 0.08$~GeV$^2$.}
  \label{fig:fm}
  \end{center}
\end{figure}

\begin{figure}[hbtp]
  \begin{center}
  \scalebox{0.6}{\includegraphics{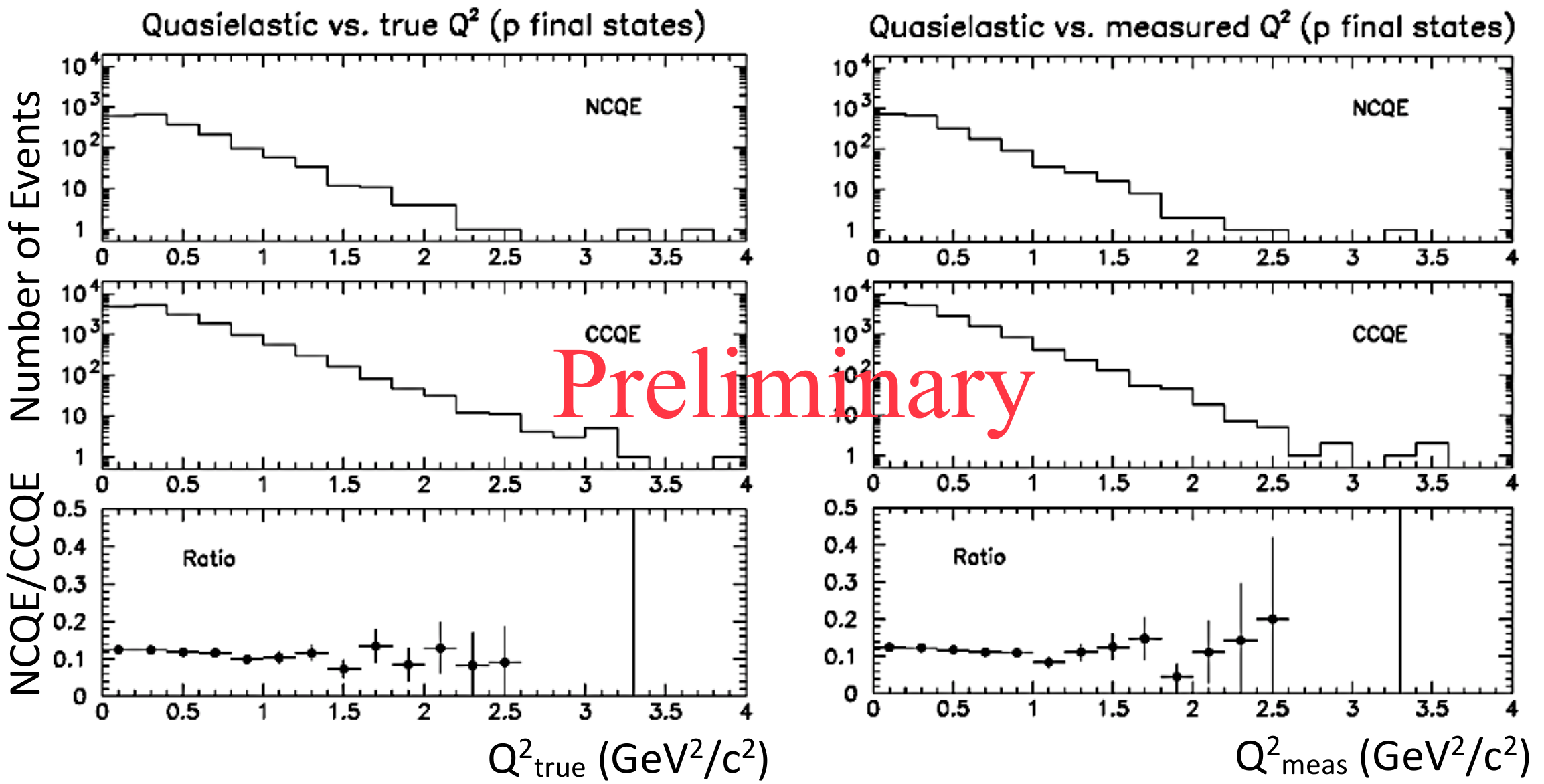}}
  \caption{Number of neutral current ($\nu p$ elastic) and charged current ($\nu n$ quasi-elastic) events as a function of $Q^2$, left for the true $Q^2$, right for the measured $Q^2$. Bottom two plots show the ratio of these two events which will be used to make measurements.}
  \label{fig:ratio}
  \end{center}
\end{figure}

The simulation only models the interaction and the Fermi motion nuclear effect, no detector effects are simulated.
Fig.~\ref{fig:fm} shows how the resolution of $Q^2$ spreads as $Q^2$ is reduced for charged current and neutral current quasi-elastic scattering.
Fortunately, by taking the ratio of neutral current to charged current events, the uncertainties due to Fermi motion are reduced.
For example, the $Q^2$ spectrum for neutral current and charged current events and their ratio is shown in Fig.~\ref{fig:ratio}.
The true $Q^2$ is on the left, the measured $Q^2$ is on the right.
Comparing the ratio at low $Q^2$ for both measured and true indicates that this nuclear effect does not prevent a good measurement at low $Q^2$.

\begin{figure}[hbtp]
  \begin{center}
  \scalebox{0.6}{\includegraphics{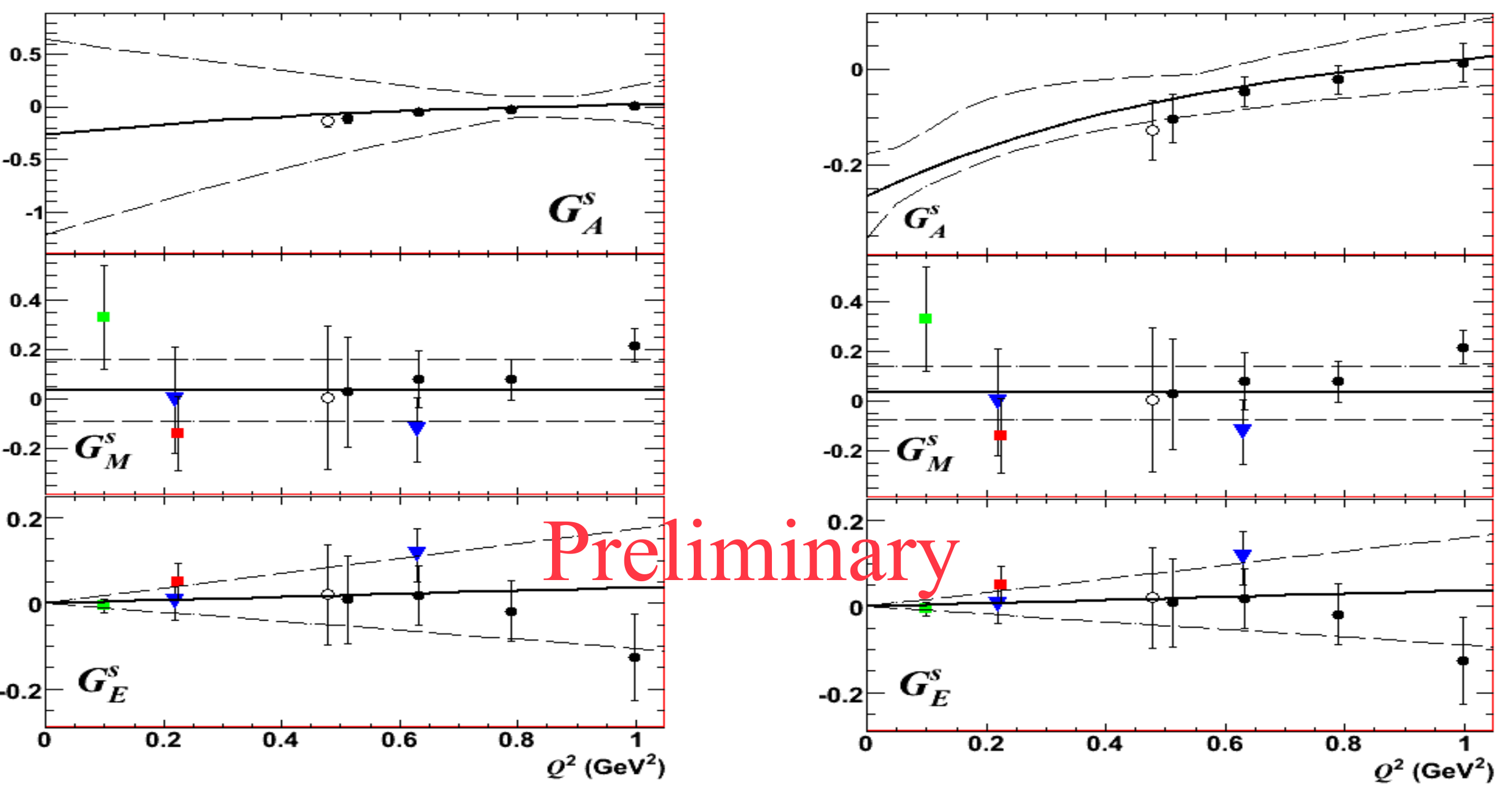}}
  \caption{Same as Fig.~\ref{fig:globalfit}, but on the right, the lines include with MicroBooNE projected data folded into the fit~\cite{SPate}\cite{SPate2}.}
  \label{fig:newff}
  \end{center}
\end{figure}

Folding the simulated MicroBooNE result into the fit, the uncertainty on the form factor $G^s_A$ is improved by an order of magnitude~\cite{SPate}\cite{SPate2}.
A comparison of the form factors before and after inclusion of the MicroBooNE simulation is shown in Fig.~\ref{fig:newff}.

\section{Experimental Considerations}
\label{sec:considerations}
As the MicroBooNE detector is commissioned, many studies will be done to understand experimental backgrounds.
The most challenging backgrounds are those anticipated from neutrons which are produced in neutrino interactions in the material around the detector.
Since they are neutral, they can mimic a neutrino interaction, and can have similar scattering topologies as neutral current events.
Neutrons are produced by neutrino interactions in the material up stream of the detector.
Low energy neutrons can bounce around the detector hall, and enter the detector in any direction, whereas fast neutrons will enter the detector's front face with neutrinos in the beam.

The low energy neutrons generally enter the sensitive area from every direction.
These will be challenging to estimate experimentally since argon does not moderate neutrons efficiently, we expect to estimate this background using Monte Carlo.

The high energy neutrons are expected to enter MicroBooNE upstream.
By counting the number of neutral current-like events within certain distances from the upstream face of the detector, a model of the number and energy of incoming neutrons can be made.
As a neutrino travels through the detector, the probability of its interacting with liquid argon is flat, it is independent of its penetration depth.
Conversely, a neutron is more likely to interact in the upstream half of the detector due to its much shorter interaction length.

If the number of high energy neutrons cannot be well modeled, auxiliary detectors may be utilized to estimate the neutron rate of the beam.
Since charged particles are often produced with the high energy neutrons, a scintillator wall in front of MicroBooNE may be used as a veto to indicate that an incoming neutron is from an upstream beam scatter.
Another option under study is to directly measure the neutron rate with a temporary low density gas TPC placed in front of MicroBooNE, which will still require a good model and simulation of neutron interactions~\cite{dctpc}.

\section{Conclusion}
\label{sec:conclusion}

The MicroBooNE neutrino detector is well equipped to measure short particle tracks as is necessary to measure protons recoiling from a neutral current interaction.
The fine resolution tracking enables detection of neutral current scatters at very low $Q^2$.
By comparing the ratio of neutral current interactions to charged current interactions in data with those in simulations, the value of the nuclear form factor $G^s_A$ can be measured at low $Q^2$.
This measurement will refine the uncertainty in the global fits of this form factor at $Q^2$ approaching zero, which provides a measurement of $\Delta s$.
The value of $\Delta s$ can modify the cross section limits of dark matter interactions by a few orders of magnitude.





\bibliographystyle{elsarticle-num}
\bibliography{taupbib.bib}







\end{document}